\documentstyle[epsfig,array,longtable]{aipproc}
\begin{document}
 
\title{Configuration Dependence of Physical Properties of a Ferroelectric
Solid Solution\thanks{To be published in the
proceedings of the Fifth Williamsburg Workshop on First-Principles Calculations
for Ferroelectrics, February 1998.}}
\author{Eric Cockayne and Karin M. Rabe}
\address{Department of Applied Physics, Yale University, P.O. Box
208284, New Haven, CT 06520-8284} 

\maketitle

\begin{abstract}
In this article, we motivate the detailed comparison of the
physical properties of individual configurations of a 
ferroelectric solid solution as a means toward developing
first principles models for these systems.
We compare energies, dielectric constants $\epsilon_{\infty}$,
mode effective charges of local polar distortions, 
and the zero temperature piezoelectric
behavior of several ordered Pb$_3$GeTe$_4$ supercells.
Cluster expansions of these properties show the
importance of second-neighbor effects, which can be
related to symmetry-preserving 
relaxation and its effect on the symmetry breaking
polar instabilities.

\end{abstract}

\section*{Introduction}

  Ferroelectric solid solutions are of great technological
importance.  For example, the largest piezoelectric response
are found in mixed ferroelectrics, 
such as Pb(Zr$_{1-x}$Ti$_x$)O$_3$.  Recently, 
giant piezoelectricity was discovered in the relaxor
ferroelectric systems 
Pb(A$_{1/3}$Nb$_{2/3}$)O$_3$--PbTiO$_3$, (A = Zn, Mg).\cite{crPar96}
Understanding the physics of ferroelectric solid solutions  
on the microscopic level
would be of great 
theoretical
interest and could also point to new ways to tune their
piezoelectric response and other
properties.

    Ab initio calculations have proved successful in relating
properties of stoichiometric ferroelectrics to phenomena on the atomic
level\cite{crCoh92}.  
Structural parameters\cite{crCoh92,crSin95},
dielectric constants, effective charges,
phonon dispersion relations\cite{crBar87,crGon92},
and polarizations\cite{crKin93}
have been
calculated from first principles.
To predict the behavior at finite temperature,
first-principles models have been 
constructed.\cite{crRab87,crZho94bt,crRab95pt,crCoc97,crKra97}.  
Generally, these models are based on a vector
representation of the local polar distortions responsible for the
ferroelectric phase transition.  
An electric dipole moment is
associated with each local distortion.  
At long distance, the
interaction between local distortions is dipole-dipole; significant
corrections appear at short range.  
Anharmonic terms, elastic
constants and strain coupling to local distortion also
appear, determining the ground state and the nature of the phase
transitions in the models.\cite{crWag97sc}  The models obtained allow for simulation
of ferroelectric phase transitions, where the Curie point is generally
in good agreement with experiment.  The models also allow
piezoelectricity and dielectric functions to be computed\cite{crGar97,crRab98}.
In the following, we discuss the modeling of ferroelectric
and piezoelectric behavior in solid solutions, and lay part of the
groundwork for a Pb$_{1-x}$Ge$_x$Te model by exploring the 
configuration dependence of certain quantities essential for 
constructing the model.

\section*{Why Compare Configurations?}

 Consider a ferroelectric solid solution.  The physical properties 
of this solid solution are the ensemble averages of
the properties of the individual ensemble members.
Thus one needs to be able to compute these physical properties
for individual ensemble members.  A given ensemble member
(configuration) can be treated as if it were an actual crystal
structure and modeled in the same manner as for a stoichiometric
ferroelectric.  

  While models for a few
chosen small unit cell configurations are obtained using this 
approach, some
model for the general configuration is necessary
in order to properly take ensemble averages.
Given the nature of the model described in the Introduction, we 
expect that in a model for the general configuration,
there 
will again be polar instabilities, a local basis 
for these distortions, and 
interactions between the local
distortions.
However, now the local polar distortions themselves and their
interactions will be site-dependent.  In principle, the site-dependence
of any quantity can be described by a cluster expansion.
The models obtained for the series of individual small unit
cell configurations can be used to obtain the cluster expansions
for a model that is valid for all configurations.

  In modeling stoichiometric ferroelectrics, two kinds of truncation
in the models are necessary to prevent an explosion of terms:  
the order of the expansion of the energy in powers of
the local distortion variable and the range of the local interactions.
The cluster expansion approach necessitates a third kind of
physically motivated truncation: the range of the cluster expansion.
Where possible, terms should be found whose configuration dependence
is unimportant and then kept constant for all configurations.
For those terms where a cluster expansion is
necessary, it should be truncated at the shortest possible range.
Toward this end, it is thus very important to establish and understand 
the configuration dependence of those quantities 
that determine the model parameters.

\begin{figure}[b!] 
\centerline{\epsfig{file=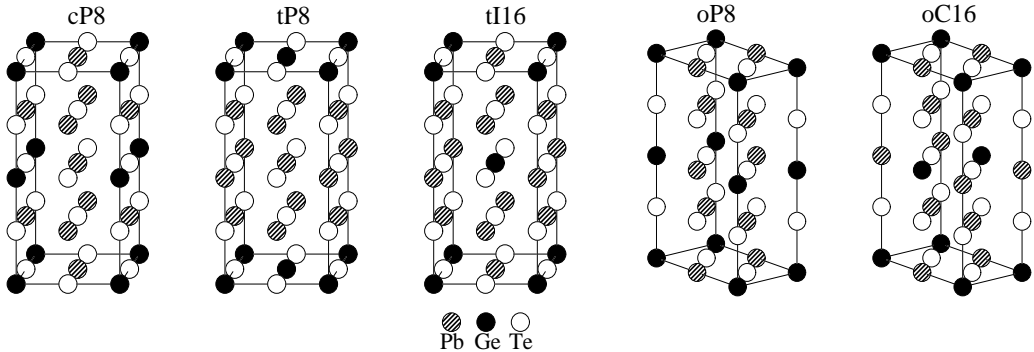,height=2.125in,width=2.5in}}
\vspace{10pt}
\caption{Pb$_3$GeTe$_4$ configurations investigated.  Each has 8 atoms
per primitive unit cell. (The Pearson symbols specify the number of
atoms in the conventional unit cell.)}
\label{crcells.fig}
\end{figure}

Pb$_{1-x}$Ge$_x$Te provides an excellent
prototype system for investigating the form of first-principles models
in solid solutions using the approach just outlined.
For all compositions above a critical 
composition $x \approx 0.005$\cite{crTak79},
Pb$_{1-x}$Ge$_x$Te undergoes a transition from a cubic
phase for $T > T_c(x)$ to a rhombohedral phase at
$T < T_c(x)$\cite{crHoh72}.
Because the endmembers of the solid solution
series have the rocksalt structure with only 2 atoms per 
primitive cell, a large number of different mixed configurations
can be investigated without the need for excessively large supercells.
 To simplify the problem further, we focus on
configuration dependence at
a fixed composition,
Pb$_{0.75}$Ge$_{0.25}$Te.  This composition was chosen to be near
the low Ge concentration regime of physical interest, while allowing
for a variety of supercells with only 8 atoms.  We present results
on the five 8-atom supercells of highest symmetry, shown in 
Figure~\ref{crcells.fig}.  In each 
configuration studied, all Ge ions are translationally equivalent
and therefore have the
same environment, greatly simplifying the analysis.  

\section*{First Principles Methods}

In our study of the structural phase transition in the
cP8 configuration\cite{crCoc97}, we give full
details of the first principles calculations.
Briefly,
we performed all calculations using density functional
theory within the local density approximation (LDA).
Ab initio pseudopotentials were used for the ions and a plane
wave basis set with cutoff energy 300 eV 
was used for the Kohn-Sham eigenfunctions.
Total energies were 
calculated via conjugate gradients optimization using
the CASTEP~2.1 package\cite{crCAS91}.  The CASTEP~2.1
package was modified\cite{crWag96} to do variational linear
response\cite{crGon92} calculations of force constants,
Born effective charges and dielectric constants.
Brillouin zone averages for the cP8 configuration were performed
using a $4 \times 4 \times 4$
Monkhorst-Pack set.  
For the other configurations, the same ${\bf k}$
point grid was used, folded into the corresponding Brillouin zone.

\section*{Symmetry Preserving Relaxation}

  We found two distinct ways in which the total energies
of the configurations that we studied could be lowered
with respect to the energy of the structures with all atoms
fixed on rocksalt positions: symmetry-preserving relaxation and
symmetry-breaking polar instabilities.   The existence of
symmetry-preserving relaxation follows from group theory.
Except for the cP8 configuration, all of the configurations
have one or more identity irreducible 
representations (irreps) among its normal modes.
An identity irrep implies an energy term that is linear in
the corresponding mode and thus a lowering of the energy
by relaxation in the ionic displacement subspace spanned
by these modes.  This relaxation neither lowers the symmetry
nor causes net polarization in the configurations studied.
The other normal modes in the systems have non-identity
irreps; thus to lowest order, the total energy is quadratic in these modes.
The existence of instabilities (normal
modes whose harmonic coefficient is negative) with their
concurrent symmetry breaking is a system-specific phenomenon, 
as is the polar nature of these instabilities.  This section
focuses on symmetry-preserving relaxation; the next section
focuses on symmetry-breaking polar instabilities.

  Before investigating symmetry-preserving relaxation,
we relaxed each configuration of Figure~\ref{crcells.fig} 
with respect to strain, holding all 
ions fixed at ideal rocksalt positions.  
At this point, the total energies all differed by less than
10 meV/(8 atom cell) and
the (pseudocubic) lattice parameters all differed by less than
0.06\%. Had all subsequent calculations been performed with respect
to the same pseudocubic lattice parameters, 
{\it e.~g.} the cP8 value of $a$ = 6.275~\AA,
the strain energies
involved would been negligible.

  Each configuration was allowed to undergo symmetry preserving
relaxation, with its strain held fixed.
The relaxed atomic coordinates can be simply described
(to within 0.03~\AA) by displacement of the 
Te ions in each Pb-Te-Ge chain
segment 0.12~\AA~toward to the Ge ion, with all
other ions held fixed.\cite{crfoot1}  Qualitatively, 
relaxation in
Pb$_{1-x}$Ge$_{x}$Te 
can be regarded as the result of
the size
mismatch between the smaller Ge and the larger Pb ions.
The energy of the cP8 configuration,
for which no symmetry-preserving relaxation occurs,
was 171 meV/cell higher than for the relaxed 
oC16 configuration, in which every Te is 
displaced.

  The energies of the relaxed configurations were fit
to a pair-only cluster expansion out to fourth cation-cation
neighbors.  In meV/cell, the total energy of the
configurations tested is given by:
\begin{equation}
E = E_0 - 2.9 \overline{N}_1 + 26.4 \overline{N}_2 
- 1.5 \overline{N}_3 + 1.3 \overline{N}_4,
\label{crrelax.eq}
\end{equation}
where $\overline{N}_i$ is the average number of $i$'th neighbor
Ge ions per Ge ion.
The cation sublattice in Pb$_{1-x}$GeTe is fcc; each
cation has 12 first neighbors cations at $(a/2,a/2,0)$, etc.,
6 second neighbors at $(0,0,a)$, etc., 24 third
neighbors at $(a/2,a/2,a)$, etc., and 12 fourth neighbors
at $(a,a,0)$, etc., where $a$ is the conventional fcc
lattice parameter.  
The sensitivity of energy to $\overline{N}_2$
is an order of magnitude larger than for the other terms, 
illustrating
the importance of the above-mentioned relaxation. 
The Ge and the Pb in a linear Ge-Te-Pb segment where the
Te undergoes significant relaxation are second
neighbors; the more Ge-Te-Pb segments, the fewer Ge-Te-Ge
segments, the lower $\overline{N}_2$, and the lower the
relaxed energy.

  The expression~\ref{crrelax.eq} is a fit of 5 relaxed cell 
energies to five parameters.  It is also important to get
an estimate of the predictive power of a cluster expansion
truncated at a smaller number of terms.  We did a series
of cluster expansions in terms of $\overline{N}_1$,
$\overline{N}_2$ and $\overline{N}_3$ to each subset of
4 configurations\cite{crfoot2} in Figure~\ref{crcells.fig}, and used
the resulting expression to calculate the energy of the
configuration that was left out.  The rms error was 
16 meV/cell and the maximum error was 20 meV/cell.  This
compares with the root mean square variance of 57 meV/cell for
the relaxed energies of the set of configurations studied,
showing modest but not excellent predictability for a
cluster expansion out to third neighbor.
In what follows, we present results based on three or
fewer configurations and the cluster expansions are truncated
where the number of parameters matches the number of unknowns.
Calculations on additional configurations would be necessary
to make any meaningful test of predictability.

\section*{Symmetry Breaking Instabilities}

  Next, we looked at energy lowering via symmetry-breaking
distortions.
  For each configuration, we then calculated the normal modes
at the zone center.  
We performed linear response force constant calculations
at ${\bf q} = 0$ for each relaxed configuration and then
diagonalized the corresponding dynamical matrix to obtain
the normal modes.
For each configuration, there were symmetry-breaking
polar instabilities, with a three-
dimensional 
vector representation centered on the Ge ion.\cite{crfoot3}
Further ${\bf q} \neq 0$ calculations on cP8 and tP8 showed
that instabilities throughout the Brillouin zone could be well-
represented by a local (lattice Wannier function\cite{crRab95lwf}) 
basis with one vector on each Ge ion.  This supports the Ge off-centering
model for ferroelectricity in 
Pb$_{1-x}$Ge$_x$Te\cite{crLog77} and gives
strong evidence that the correct form for a first-principles
model for arbitrary configurations will have one vector per Ge ion. 

  In order to appropriately model the long-range physics
within our models, both the electronic dielectric tensor
${\bf \epsilon}_{\infty}$ and the polarization associated
with the local instabilities must be determined.
These are linear response functions of the relaxed 
high-symmetry structures.
We have calculated the dielectric tensors ${\bf \epsilon}_{\infty}$
and the Born effective charges
for the cP8, tP8, and tI16 configurations.
The Born effective charge tensors ${\bf Z}^{\star}_j$ and the
normal mode ionic displacement pattern ${\bf u}_{j\beta}$
associated with the zone-center instability which transforms
like the vector component $\alpha$ were used to calculate the
so-called mode effective charge tensor through
\begin{equation}
\overline{Z}^{\star}_{\alpha\beta} = 
\sum_{j\gamma} (Z^{\star}_j)_{\alpha\gamma} (u_{j\beta})_{\gamma}
\label{creqmec}
\end{equation}
The normal modes were all normalized such that
$\sum_j |{\bf u}_{j\beta}|^2$ was 1 \AA$^2$.

  The results for the 
${\bf \epsilon}_{\infty}$ and $\overline Z^{\star}$ tensors
are shown in Table~\ref{crez.tbl}.  Fitting to a pairwise
cluster expansion out to second neighbor, we obtain:
\begin{equation}
(\epsilon_{\infty})_{\alpha\alpha} =
42.5 - 0.175 \overline{N}_{1\parallel} - 
0.1 \overline{N}_{1\perp} + 1.7 \overline{N}_{2\parallel}
+ 0.2 \overline{N}_{2\perp},
\end{equation}
where $\overline{N}_{i\perp}$ is the mean number of $i$'th
neighbor Ge ions per Ge ion in a direction perpendicular
to $\alpha$ and $\overline{N}_{i\parallel}$ is the mean
number of $i$'th neighbor Ge ions per Ge ion in a
direction having nonzero component along $\alpha$.
Likewise, the mode effective charge is given (in units of e\AA) 
by
\begin{equation}
\overline{Z}^{\star}_{\alpha\alpha} =
17.31 -0.14 \overline{N}_{1\parallel} -     
0.12 \overline{N}_{1\perp} - 2.17 \overline{N}_{2\parallel}
- 0.22 \overline{N}_{2\perp}.
\end{equation}
The relative configuration dependence of $\overline{Z}^{\star}$ is
larger than that of $\epsilon_{\infty}$.  The three
configurations all have diagonal
dielectric and mode effective charge tensors;
it should be noted that off-diagonal terms
will be nonzero in the general (asymmetric) configuration.
For the mode effective charge tensors,
the more Ge-Te-Pb or Pb-Te-Ge segments along $z$, the
lower $\overline{N}_{2\parallel}$ and the 
higher $\overline{Z}^{\star}_{zz}$.
We have previously
shown how this increase can be explained in terms of local
bonding\cite{crCoc98}.
The differences in the Born effective charges of 
individual ions between configurations is relatively
small; 
it is primarily the difference in the normal mode 
eigenvector
between configurations
that leads to the increase
in $\overline{Z}^{\star}$.
In particular, in Ge-Te-Pb chain segments, relaxation of the
Te away from the Pb leads effectively to an 
off-centering instability of the Pb ion that does not
occur when the Pb-Te distance is smaller.\cite{crCoc98}
The active participation of the Pb ion helps to
increase $\overline{Z}^{\star}$.

\begin{table}[h]
\begin{center}
\caption{Electronic dielectric constant and mode effective
charges in three Pb$_3$GeTe$_4$ configurations.}
\begin{tabular}{||c|c|c|c||}
\hline  Component  & cP8  & tP8 & tI16 \\
\hline $(\epsilon_{\infty})_{xx}$  & 46.7  & 45.6   & 46.3 \\
\hline $(\epsilon_{\infty})_{yy}$ &  46.7  & 45.6   & 46.3 \\
\hline $(\epsilon_{\infty})_{zz}$ &  46.7  & 42.9   & 43.3 \\
\hline $\overline{Z}^{\star}_{xx}$ &  12.10  & 11.97 & 12.54   \\
\hline $\overline{Z}^{\star}_{yy}$ &  12.10  & 11.97 & 12.54   \\
\hline $\overline{Z}^{\star}_{zz}$ &  12.10  & 15.95 & 16.44  \\
\hline
\end{tabular}
\label{crez.tbl}
\end{center}
\end{table}

\section*{Piezoelectricity at Zero Temperature}

 The above calculations were for relaxed high symmetry states.
We now turn our attention to properties of the fully distorted
LDA ground states.
  We have previously reported the piezoelectric tensors at
zero temperature for the cP8 and tP8 configurations\cite{crCoc98}.
In this section, we give further results for the tI16 configuration,
write a simple cluster expansion for the zero-temperature 
piezoelectricity
in Pb$_{0.75}$Ge$_{0.25}$Te, and use this expansion to estimate
the piezoelectric tensor of the disordered system.

\begin{table}[h]
\begin{center}
\caption{LDA ground state of Pb$_3$GeTe$_4$-tI16.
Monoclinic, space group Cm, $a$ = 8.984~\AA,
$b$ = 8.876~\AA, $c$ = 7.738~\AA,
$\beta$ = 54.08$^{\circ}$.}
\begin{tabular}{||c|c|c|c|c||}
\hline Atom  &  Wyckoff position  & x   & y & z \\
\hline Pb & 2(a) & 0.5034 & 0 & 0.0061 \\
\hline Pb & 4(b) & 0.5020 & 0.2489 & 0.5000 \\
\hline Ge & 2(a) & 0.0291 & 0 & 0.0069 \\
\hline Te & 2(a) & 0.2355 & 0 & 0.5158 \\
\hline Te & 2(a) & 0.7578 & 0 & 0.4753 \\
\hline Te & 4(b) & 0.2391 & 0.2383 & 0.9976 \\
\hline
\end{tabular}
\label{crti16g.tbl}
\end{center}
\end{table}

The LDA ground state for the tI16 configuration is given in
Table~\ref{crti16g.tbl}.
In Table~\ref{crpiezo.tbl}, we give the full piezoelectric tensors
for the LDA ground states of the cP8, tP8 and tI16 configurations.  
In each case, the orientation is with respect to the axes of 
Figure~\ref{crcells.fig}, and the symmetry has been broken such that the
Ge displacement and the polarization lie in the $+\{111\}$
quadrant.

\begin{table}[h]
\begin{center}
\caption{Comparison of piezoelectric strain tensors for
the ground states of three Pb$_3$GeTe$_4$
configurations
(in $C/m^2$).  Components in parentheses are equal to other
components via symmetry.
Components which do not appear in the table are related
by symmetry to those that do; {\it e.g.} $e_{22} = e_{11}$
for each configuration}
\begin{tabular}{||c|c|c|c||}
\hline Component  &  cP8  & tP8   & tI16 \\
\hline $e_{11}$ &   1.8  &  2.5 & 2.4 \\
\hline $e_{12}$ &  -0.6  & -1.2 & -1.2 \\
\hline $e_{13}$ & (-0.6) & -0.9 & -0.6 \\
\hline $e_{14}$ &  -0.2  & -0.6 & -0.5 \\
\hline $e_{15}$ &   1.0  &  1.2 & 0.2 \\
\hline $e_{16}$ &  (1.0) &  1.1 & 1.1 \\
\hline $e_{31}$ & (-0.6) & -0.3 & -0.5 \\
\hline $e_{33}$ &  (1.8) &  5.1 & 6.5 \\
\hline $e_{34}$ &  (1.0) &  2.2 & 6.4 \\
\hline $e_{36}$ & (-0.2) & -0.5 & -0.7 \\
\hline
\end{tabular}
\label{crpiezo.tbl}
\end{center}
\end{table}

  If only pair terms are included in the cluster expansion for 
piezoelectricity, then the piezoelectric tensor is given by
\begin{equation}
{\bf e} = {\bf e}_0 +  \sum_{\bf d} \overline{N}_{\bf d} {\bf e}_{\bf d},
\label{crpclust.eq}
\end{equation}
where ${\bf e}$ is the piezoelectric tensor, ${\bf e}_0$ is
a configuration-independent term,
$\{{\bf d}\}$ is the set of cation-cation
separations, $\overline{N}_{\bf d}$ is the average number of Ge 
neighbors per Ge ion at separation ${\bf d}$, and ${\bf e}_{\bf d}$ is
the correction to the piezoelectric tensor due to neighbors
at separation ${\bf d}$.  
We are now dealing with a property of 
Pb$_{0.75}$Ge$_{0.25}$Te at zero temperature;
the discussion of piezoelectric tensor
in this section only applies to the subensemble of zero
temperature structures that have been poled by a field in
the $+ {111}$ direction.  The subensemble has rhombohedral
symmetry; the average piezoelectric tensor has only four
independent components: $e_{11}$, $e_{12}$, $e_{14}$ and
$e_{15}$.

  Symmetry constrains the form of the individual
${\bf e}_{\bf d}$.  Consider for example the effect of
adding one Ge-Ge pair separated by $(a/2,a/2,0)$ to a
perfectly rhombohedral configuration.  The global symmetry is
broken to monoclinic.  By writing the appropriate symmetrized
form of the piezoelectric tensor for a monoclinic system
and subtracting that for a rhombohedral system, the
correct symmetry for ${\bf e}_{(a/2,a/2,0)}$ is obtained.
It is straightforward to show that
the tensors ${\bf e}_{\bf d}$ for two different
${\bf d}$ are related if and only if the two ${\bf d}$ are equivalent
by symmetry under the rhombohedral group.
In the cation sublattice, the 12 first neighbors break
under rhombohedral distortion
into 2 groups of 6, while the 6 second neighbors are all
equivalent.

 The cP8, tP8 and tI16 configurations are sufficient to
separate the first neighbor and second neighbor contributions
to the piezoelectric tensor, but insufficient for determining
the independent contributions of the two kinds of first
neighbor.  In Table~\ref{crpclust.tbl}, we give ${\bf e}_0$ and the 
contributions 
${\bf e}_{(a/2,a/2,0)} + {\bf e}_{(a/2,-a/2,0)}$ and
${\bf e}_{(0,0,a)}$.  
Both the tensors 
${\bf e}_{(a/2,a/2,0)} + {\bf e}_{(a/2,-a/2,0)}$ and
${\bf e}_{(0,0,a)}$
have the same symmetry as that for the tP8 and tI16
ground state piezoelectric tensors.

The dominant terms reflect the
piezoelectric components that change the most from
configuration to configuration.  The lack of second neighbor Ge
pairs along (0,0,a) tends to increase the value of
$e_{33}$, while the lack of Ge pairs separated
by $(\pm a/2, \pm a/2, 0)$ and $(0,0,a)$ both increase
the value of $e_{34}$.  The physics of the $(0,0,a)$
pairs is clear: the presence of a Ge-Te-Pb chain
segment 
means that there is relaxation of the Te atom
joining them and thus a weakening of the instability
that transforms like $z$.
A weak instability implies
large response\cite{crCoc98,crRab98}.  The $e_{33}$ and $e_{34}$
components are the ones enhanced because the instability
along $\hat{z}$ is effectively
coupled most strongly to these components.\cite{crCoc97}.
The exact source
of the large influence of first neighbor pairs on 
$e_{34}$ has yet to be determined.

\begin{table}[h]
\begin{center}
\caption{Terms in cluster expansion of 
Pb$_{0.75}$Ge$_{0.25}$Te piezoelectric tensor (in $C/m^2$)}
\begin{tabular}{||c|c|c|c||}
\hline  Component  & ${\bf e}_0$    & ${\bf e}_{(a/2,a/2,0)} + {\bf e}_{(a/2,-a/2,0)}$ 
  & ${\bf e}_{(0,0,a)}$ \\
\hline $e_{11}$ &   7.7  &  0.1 & -0.3 \\
\hline $e_{12}$ &  -1.0  &  0.2 &  0.3 \\
\hline $e_{13}$ & (-1.0) & -0.1 &  0.0 \\
\hline $e_{14}$ &  -1.2  &  0.0 &  0.1 \\
\hline $e_{15}$ &   5.6  &  0.5 &  0.4 \\
\hline $e_{16}$ &  (5.6) &  0.0 &  0.0 \\
\hline $e_{31}$ & (-1.0) &  0.1 & -0.1 \\
\hline $e_{33}$ &  (7.7) & -0.7 & -2.4 \\
\hline $e_{34}$ &  (5.6) & -2.2 & -2.7 \\
\hline $e_{36}$ & (-1.2) &  0.1 &  0.2 \\
\hline
\end{tabular}
\label{crpclust.tbl}
\end{center}
\end{table}

  Given the data in Table~\ref{crpclust.tbl}, 
it is possible to estimate the piezoelectric tensor of 
disordered Pb$_{0.75}$Ge$_{0.25}$Te.  We assume
for present purposes that
all members of the poled subensemble are equiprobable.  
Then there are no spatial correlations and
$\overline{N}_{\bf d} = 0.25$ for all ${\bf d}$.
Using the values in 
Table~\ref{crpclust.tbl} and Eq.~\ref{crpclust.eq},
we estimate for disordered Pb$_{0.75}$Ge$_{0.25}$Te that
$e_{11} = 5.9$, $e_{12} = -0.9$, $e_{14} = -1.0$, and
$e_{15} = 3.7$,
reproducing the tensor form expected for
rhombohedral symmetry.
The estimated values of the components suggest that the 
piezoelectric response of the cP8 configuration is
particularly
unrepresentative for the disordered system.   
Because of the strong dependence of piezoelectric
response on the magnitude of instability in a system
and the configuration dependence of the magnitude
of local instability,
piezoelectric results on single
high symmetry supercells of mixed ferroelectrics should 
be treated with caution.

\section*{Conclusions}

 This work describes the necessity of cluster expansions
in first principles models for 
piezoelectricity and ferroelectricity in solid
solutions and the importance of comparing the properties
of different configurations.
In the specific example of 
Pb$_{0.75}$Ge$_{0.25}$Te, strong configuration dependence
is found for relaxed energies, the nature of the local
polar instabilities, and the polarization associated with
these instabilities.  A good model for ferroelectricity in
a mixed system must be able to account for this
variability.  
We show how results of the zero temperature piezoelectricity
on several ordered supercells can be used to obtain an
estimate of the zero temperature piezoelectricity of a
completely disordered cell.

\section*{Acknowledgements}

This work was supported by ONR N00014-97-J-0047.


\begin{references}

\bibitem{crPar96} S.-E. Park and T. R. Strout, in
{\it 1996 IEEE Ultrasonics Symposium Proceedings}
(New York: IEEE, 1996), v.2, p. 935;
{\it J. Appl. Phys.} {\bf 82}, 1804 (1997), and
references therein.

\bibitem{crCoh92} R. E. Cohen, {\it Nature}
{\bf 358}, 136 (1992).

\bibitem{crSin95} D. J. Singh, {\it Phys. Rev. B}
{\bf 52}, 12559 (1995).

\bibitem{crBar87} S. Baroni, P. Giannozzi and A. Testa,
{\it Phys. Rev. Lett.} {\bf 58}, 1861 (1987).


\bibitem{crGon92} X. Gonze, D. C. Allan and M. P. Teter,
{\it Phys. Rev. Lett.} {\bf 68}, 3603 (1992).

\bibitem{crKin93} R. D. King-Smith and D. Vanderbilt,
{\it Phys. Rev. B} {\bf 47}, 1651 (1993).

\bibitem{crRab87} K. M. Rabe and J. D. Joannopoulos,
{\it Phys. Rev. Lett.} {\bf 59}, 570 (1987).

\bibitem{crZho94bt} W. Zhong, D. Vanderbilt, and K. M. Rabe,
{\it Phys. Rev. Lett.} {\bf 73}, 1861 (1994);
{\it Phys. Rev. B} {\bf 52}, 6301 (1995).

\bibitem{crRab95pt} K. M. Rabe and U. V. Waghmare, {\it Ferroelectrics}
{\bf 164}, 15 (1995); U. V. Waghmare and K. M. Rabe,
{\it Phys. Rev. B} {\bf 55}, 6161 (1997).

\bibitem{crCoc97} E. Cockayne and K. M. Rabe, {\it Phys. Rev. B}
{\bf 56}, 7947 (1997).

\bibitem{crKra97} H. Krakauer, R. Yu, C.-Z. Wang, K. M. Rabe and U. V.
Waghmare, unpublished (cond-mat/9710088); 
H, Krakauer, R. Yu, C. Z. Wang, and C. Lasota, 
{\it Ferroelectrics} (to be published).

\bibitem{crWag97sc} K. M. Rabe and U. V. Waghmare, {\it Phil. Trans.
R. Soc. Lond.} {\bf A 354}, 2897 (1996).

\bibitem{crGar97} A. Garcia and D. Vanderbilt, unpublished 
(cond-mat/9712312).

\bibitem{crRab98} K. M. Rabe, these Proceedings.

\bibitem{crTak79} S. Takaoka and K. Murase,
{\it Phys. Rev. B} {\bf 20}, 2823 (1979).

\bibitem{crHoh72} D. K. Hohnke, H. Holloway and S. Kaiser,
{\it J. Phys. Chem. Solids} {\bf 33}, 2053 (1972).

\bibitem{crCAS91} M. C. Payne et al., ``CASTEP~2.1",
Cavendish Laboratory, University of Cambridge (1991).

\bibitem{crWag96} U. Waghmare, Ph.~D. thesis, Yale University
(1996).

\bibitem{crfoot1} If a Te atom belongs to Pb-Te-Ge chain segments
in more than one Cartesian direction, as for example the $z = 0$
Te atoms in the oP8 and oC16 configurations, then its relaxation
is given approximately by the vector sum of the appopriate
0.12~\AA~relaxations for the individual chains.

\bibitem{crfoot2} For technical reasons, the subset of four
configurations with cP8 missing could not be used in this
analysis.  The particular values of $\overline{N}_1$, 
$\overline{N}_2$, and $\overline{N}_3$ for that set of
four configurations gives an indeterminate set of linear
equations.

\bibitem{crfoot3} The vector representation breaks into 
irreps according to the point group of the
Ge site in the given configuration.

\bibitem{crLog77} Yu. A. Logachev and B. Ya. Mo\u izhes,
{\it Sov. Phys. Solid State} {\bf 19}, 1635 (1977).
({\it Fiz. Tverd. Tela} (Leningrad) {\bf 19}, 2793 (1977)).

\bibitem{crCoc98} E. Cockayne and K. M. Rabe, unpublished
(cond-mat/9712232).

\bibitem{crRab95lwf} K. M. Rabe and U. V. Waghmare,
{\it Phys. Rev. B} {\bf 52}, 13236 (1995).


\end{references}
\end{document}